\documentclass[aps,twocolumn,epsf,floats,pre,nofootinbib,showkeys,showpacs]{revtex4-1}
\usepackage{graphics,graphicx,epsfig}
\usepackage{amssymb,color}
\usepackage{epsf,epstopdf,wrapfig}
\usepackage{amsmath}
\usepackage{amsmath}
\usepackage{amssymb}
\usepackage{graphicx}
\usepackage{bm}
\usepackage{xr}
\usepackage{scrextend}
\begin{document}


\title{Temperature dependence of the surface plasmon resonance in small electron gas fragments, self consistent field approximation.}

\author{C. Fasolato$^{1}$,
F. Sacchetti$^{2}$,
P. Tozzi$^{3}$,
C. Petrillo$^{2}$
}
\affiliation{$^1$Dipartimento di Fisica, Universit\`a di Roma Sapienza, Italy}
\affiliation{$^2$Dipartimento di Fisica, Universit\'a di Perugia and Istituto per l'Officina Materiali of CNR, Via A. Pascoli, I-06123 Perugia, Italy}
\affiliation{$^3$Dipartimento di Fisica, Universit\`a di Perugia, Via Alessandro Pascoli, I-06123 Perugia, Italy}

\begin{abstract}

The temperature dependence of the surface plasmon resonance in small metal spheres is calculated using an electron gas model within the Random Phase Approximation. The calculation is mainly devoted to the study of spheres with diameters up to at least 10 nm, where quantum effects can still be relevant and simple plasmon pole approximation for the dielectric function is no more appropriate. We find a possible blue shift of the plasmon resonance position when the temperature is increased while keeping the size of the sphere fixed. The blue shift is appreciable only when the temperature is a large fraction of the Fermi energy. These results provide a guide for pump and probe experiments with a high time resolution, and tailored to study the excited electron system before thermalisation with the lattice takes place.

\end{abstract}

\maketitle

\section{Introduction}

The electron gas is a proven prototype for interacting electrons in metals and it is a powerful reference model in the development of modern theories of many electron systems. The knowledge of the electron gas as a highly symmetric model containing, nonetheless, the physics of the real systems, has been instrumental to the development of the Density Functional Theory \cite{kohn} (DFT). So, there exists an enormous body of studies \cite{mahan} on the electron gas properties, even though a less touched subject concerns the properties of the electron gas when confined within a limited volume \cite{simgas,sfere1,sfere2,sfere3}.

Small metal fragments have been extensively investigated over the last years because of their very exciting optical properties that are exploited for applications in different areas \cite{rev16,qm1,qm2,qm3}. Small, that is nanometer scale, metal spheres are used in optical techniques, like SERS, SEIRA, etc., whose empirical development relies upon the classical Mie's theory \cite{mie} of light diffusion from a metal surface, which provides a reasonable working model for experiments in the visible region \cite{claudia}. In practical applications rather small nanoparticles are employed, which might cause quantum mechanics effects to show up following the transition from the continuous energy spectrum of an extended system into the discrete energy spectrum of a size-limited system. Although quantum effects should be visible in optical spectra, no experimental evidence is available. Moereover, there are controversial indications on quantum effects as being quite visible in small spheres \cite{simgas} against their conjectured suppression. At the core of the debate is the capability of a proper treatment of the dynamic screening\footnote{From the book of Pines and Nozieres \cite{pinnoz}: {\it Indeed, the central problem in developing a divergence-free theory of electron systems is that of introducing the concept of dynamic screening in consistent fashion}.} in the extended versus confined electron system.

The theoretical approach is nowadays challenged by the experimental possibilities opened by new sources like the Free Electron Lasers operating in a wide wavelength region \cite{fel}. Indeed, it is becoming possible to get experimental data on the behavior of materials in conditions of a highly excited electron system, which enhances the role of the {\it dynamical} screening.

The seminal paper of Wood and Ashcroft \cite{ash} on optical response of small fragments of electron gas, initiated the theoretical research on specific fragments containing enough electrons using the most advanced techniques \cite{simgas,qm2,tddft,qm4,qm6,qm7,qm8}. These modern techniques are very efficient with a solid theoretical foundation, based on the DFT, extended to the time dependent problems \cite{tddft}, although two basic limitations affect such an approach. First, the size of the fragment is too small and second, and more important, the calculation is purely numerical so limiting the understanding of the physical mechanisms responsible for the observed behavior. Whereas the first condition might be relaxed by improving the numerical technique, the second one is intrinsic to fully numerical approaches. Further, the DFT is tailored to describe zero-temperature problems, whereas the investigation of quantum effects in small fragments finds its counterpart in the experiments on matter in extreme conditions. This possibility is offered by the high energy density available at the new coherent light sources, the Free Electron Lasers (FEL), over wide wavelength regions \cite{fel}. FELs experiments can be exploited to investigate the behavior of matter at extreme temperature conditions, as T$ > 10^4$ K. The good time resolution achievable in pump and probe experiments allows first for strongly exciting the electron system, which emphasizes the role of the dynamic screening, and, second, for probing the optical response at these conditions. The finite temperature is still a challenging problem even for the electron gas \cite{mcht,mcht2}, i.e. a system where the most advanced theoretical and simulation techniques can be applied.

In this paper we present a study of the longitudinal dielectric function of a fragment of electron gas using the self-consistent approximation \cite{cohen} for electrons confined in a finite volume $\Omega$. A spherical volume $\Omega$ is chosen because it corresponds to a more realistic shape and it is characterized by non-uniformly spaced energy levels, a condition that is relevant when searching for quantum effects evidence. This study goes beyond that presented in Ref. \cite{ash} on a fragment of electron gas treated within the Random Phase Approximation (RPA). 

The present approach enables to study a system containing several thousands of electrons to recover the surface plasmon response using the classical boundary conditions \cite{mie, ash}. In the following sections the fundamental results of the self-consistent field approach \cite{cohen} are recalled for the straightforward application to the confined electron gas, in order to identify the quantum effects related to the discrete distribution of the electron states arising from the finite volume boundary conditions.

Our calculation is computationally fast enough to enable the investigation of the effects brought about by the change of relevant parameters like the electron density $n_e = 3/4 \pi (r_s a_B)^3$ ($a_B$, Bohr radius) and the size of the fragments. We focus on the temperature dependence of the surface plasmon resonance in a single gold nanoparticle up to T $\simeq 10^5 K$, that is $k_B$ T = 1 Ryd.

\section {Calculation of the dielectric function in a finite volume}

We start from the general result of the self-consistent field approximation \cite{cohen} and consider an electron gas confined in the finite volume $\Omega$. In a finite system the boundaries break down the translational symmetry of the electron gas thus introducing several effects which are not present in the extended system \cite{ash}.

Let $\phi_k({\bf r})$ be the {\it single particle} wave function, $k$ a generic quantum number, and $\epsilon_k$ the associated energy. The actual expression of the wave function depends on the shape of the system volume and analytical forms are available for the parallelepiped and the sphere cases. For a sphere, the wave function $\phi_k({\bf r})$ is proportional to the product of a spherical Bessel function $j_l(k_n r)$ times a spherical harmonic $Y_{l m}({\bf \hat r})$. The corresponding energy $\epsilon_{n l}$ is given by the $n$-th zero $z_{n l}$ of $j_l(x)$, that is $\epsilon_{n l} = \hbar^2 k_n^2 / 2 m = \hbar^2 z_{n l}^2 / 2 m R^2$, being $R$ the sphere  radius. Unfortunately, there is no explicit formula for the values of the zeros of the spherical Bessel functions when $l > 0$, therefore they must be calculated numerically. It is important to note that, apart from the case $l = 0$, the spacing of $z_{n l}$ is not uniform and there is a continuous, almost linear, increase of the position of the first zero on increasing $l$.

The diagonal part of the longitudinal dielectric function $\epsilon(q,\omega)$, at thermodynamic equilibrium and in the frequency domain, can be obtained from the von Neumann-Liouville equation for the single particle density matrix linearized in order to derive the response function of the system to an external (weak and adiabatically switched) electric potential. We confine ourselves to systems with a paired spin distribution. The result is:

\begin{equation}
\epsilon(q,\omega) = 1 - 2 v(q) \, \lim_{\eta \to 0^+} \, \sum_{k k'} \, {f(E_{k'}) - f(E_k) \over E_{k'} - E_k - \hbar \omega + i \eta} M^2_{k k'}({\bf q})\\
\label{eq1}
\end{equation}

\noindent
where $f(E)$ is the (Fermi) population factor, $\eta$ is a vanishingly small positive number that governs the adiabatic switching of the external potential, $v(q)$ is the Fourier transform of the bare Coulomb electron potential $4 \pi e^2 / q^2$ and $M^2_{k k'}({\bf q})$ is the matrix element of the Fourier transform of the electron density. This expression reduces to the well known Lindhard dielectric function for the bulk electron gas and it does not contain further approximation other than those related to the RPA \cite{mahan, cohen}. 

We observe that in the case of a small fragment, the broken translational symmetry allows also for intra-band transitions that are forbidden in the homogeneous gas.

A reliable calculation of the dielectric function requires a highly accurate calculation of the matrix elements. This task becomes increasingly complex for larger sizes of the fragments; a situation that is relevant for the present calculation, as small fragments can be treated using atomistic approaches (e.g. TDDFT). In the case of a spherical fragment, the calculation of the matrix elements is more complex as no fully analytical formulation exists. The matrix elements are expressed in terms of integrals of three spherical Bessel functions and $3j$ symbols. Through a proper manipulation of these formulas, the calculation reduces to the sum of Bessel function integrals and $3j$ symbols that depend only on the three angular quantum numbers $l_1, l_2, l_3$ over a wide range of values, namely up to 100-200 depending on the size of the particle. Considering that the matrix elements depend on $q$ only through the integrals of the Bessel functions, the calculation of the $3j$ symbols can be carried out just once and then stored. Similarly, the matrix elements are calculated at each $q$ value and stored to perform the calculation of the dielectric function as a function of the energy. 

To determine the effect of the parameter $\eta$, we note that the limit as a function of $\eta$ cannot be carried out numerically and the only possibility is to calculate $\epsilon(q,\omega)$ at different values of $\eta$. In any case, $\eta$ should be considered with some care. Indeed, $1 \over \eta$ is often identified as a relaxation time responsible for  broadening the $quantum$ peaks that are expected to appear as a consequence of the discreteness of the electron energy spectrum in the confined system. This straightforward relaxation time approximation is, in our opinion, not correct since the continuity equation is locally violated as discussed in Ref. \cite{ash}. For a more detailed discussion on the problem of producing a conserving theoretical approach one can refer to the reference paper of Baym and Kadanoff \cite{byka} and to some applications like that of Refs. \citep{ash2,merm}. 

\section{Numerical calculation of the dielectric function}

The numerical determination of $\epsilon (q, \omega)$ requires, as said, the calculation of integrals of the spherical Bessel functions and $3j$-symbols over a wide range of angular momentum with high accuracy. In the high temperature case, the maximum angular momentum to be included is fairly large but the calculation could be carried on in a rather accurate way. For a safely accurate calculation up to temperatures T $= 10^5$ K in the case of a 10 nm sphere, we decided to include all the contributions up to $l = 200$, so that energies up to 5 Ryd were considered. Accordingly, the integrals of the Bessel functions had to be calculated with a six digit accuracy. The calculation of the $3j$ symbols was performed following the proposal of Ref. \citep{3jcal} using a specifically written code. 
After several tests, we concluded that the calculation of the dielectric function was accurate to better than three digits. A description of the numerical techniques is given in a more extended paper that is in preparation.

\begin{figure}
\centering
\includegraphics[width=8.5cm]{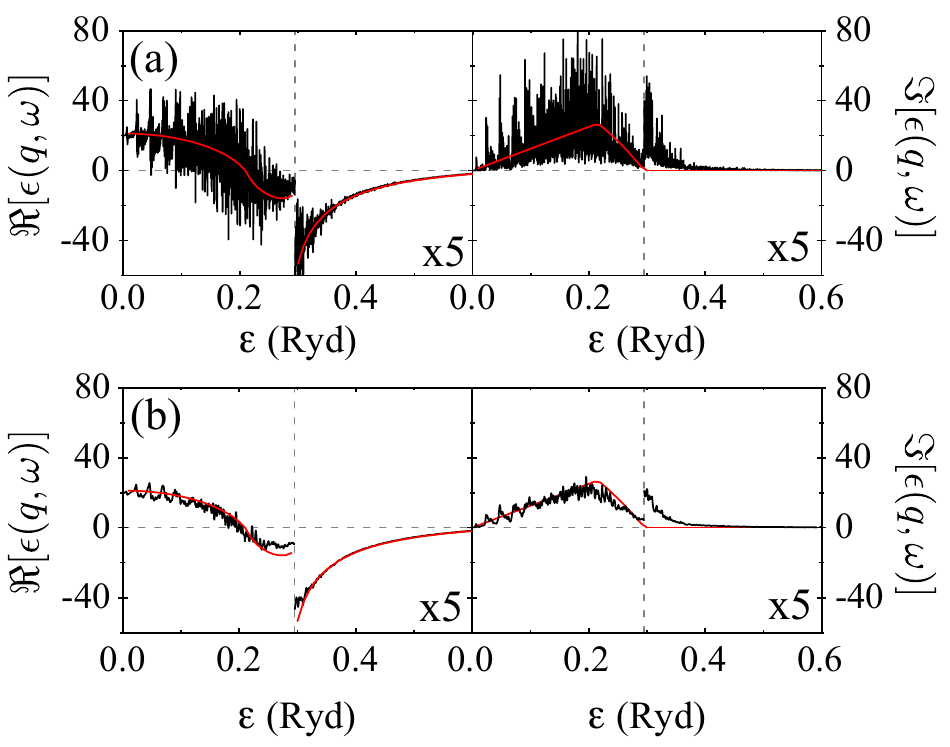}
\caption{Dielectric function of a sphere (full black line) with $r_s = 3$ and $R = 3.1$ nm, performed at $q$ = 0.2 a.u. and  T $= 0$ K. Panel (a), $\eta$ = 0.0001 Ryd, left plot, real part of $\epsilon(q,\omega)$, right plot, imaginary part of $\epsilon(q,\omega)$. The horizontal dashed line is the zero level and the vertical dashed line indicates the boundary of the particle-hole excitation region (see text) and also the position of the scale change (times 5) used to make visible the trend at high $\hbar \omega$. Panel (b), the same as Panel a but with $\eta$ = 0.001 Ryd.}
\label{fig1}
\end{figure}

The calculation of the dielectric function was performed at $r_s$ = 3 (the typical average electron density of a gold particle) for several $q$ values and for an energy range up to 1 Ryd, with varying $\eta$. As a typical result, we show in Fig.\ref{fig1} the dielectric function of a sphere with radius $R = 3.1$ nm, at $q$ = 0.2 a.u. and  T $= 0$ K.  The results obtained using $\eta = 0.0001$ Ryd and $\eta = 0.001$ Ryd are shown in the two panels of Fig.\ref{fig1}.

The overall trend is similar to the RPA result in the homogeneous electron gas, although a complex structure is apparent in the low energy region, which is clearly related to the discrete level structure of the fragment. Also, while the imaginary part of the dielectric function of the electron gas is zero outside the particle-hole pair region, limited by the two parabolic functions $\hbar \omega / (2 k_F)^2 = (q/2 k_F)^2 + q/2 k_F$ and $\hbar \omega / (2 k_F)^2 = (q/2 k_F)^2 - q/2 k_F$, the same function calculated for the sphere can extend beyond this region, due to the breaking of the translational symmetry for the reduced spherical volume. We note also that the $\eta$ = 0.001 Ryd corresponds to a relaxation time $\tau \simeq 5 \, \cdot \, 10^{-14}$ s$^{-1}$, a value fairly higher than that derived from the Drude model for metals Ref. \cite{merash}. 

Further insight on the temperature dependence of the plasmon resonance is provided by the study of the response function which is proportional to the extinction cross section, under different conditions \citep{mie}, i.e.
\begin{eqnarray}
{\cal A}(q,\omega) = -\Im \Biggl [{1 \over \epsilon(q,\omega) + C \epsilon_m} \Biggr ]
\end{eqnarray}

\noindent
where $\epsilon_m$ is the dielectric function of the external medium and $C$ is a constant equal to 0 for a bulk system, to 1 for an infinite surface and to 2 for a sphere \cite{mie}. We assumed $\epsilon_m = 1$ corresponding to the case of particles in the empty space, as done in Ref. \cite{simgas} so that a direct comparison is possible for the case of the $R = 1.34$ nm sphere. 

\begin{figure}
\centering
\includegraphics[width=8.5cm]{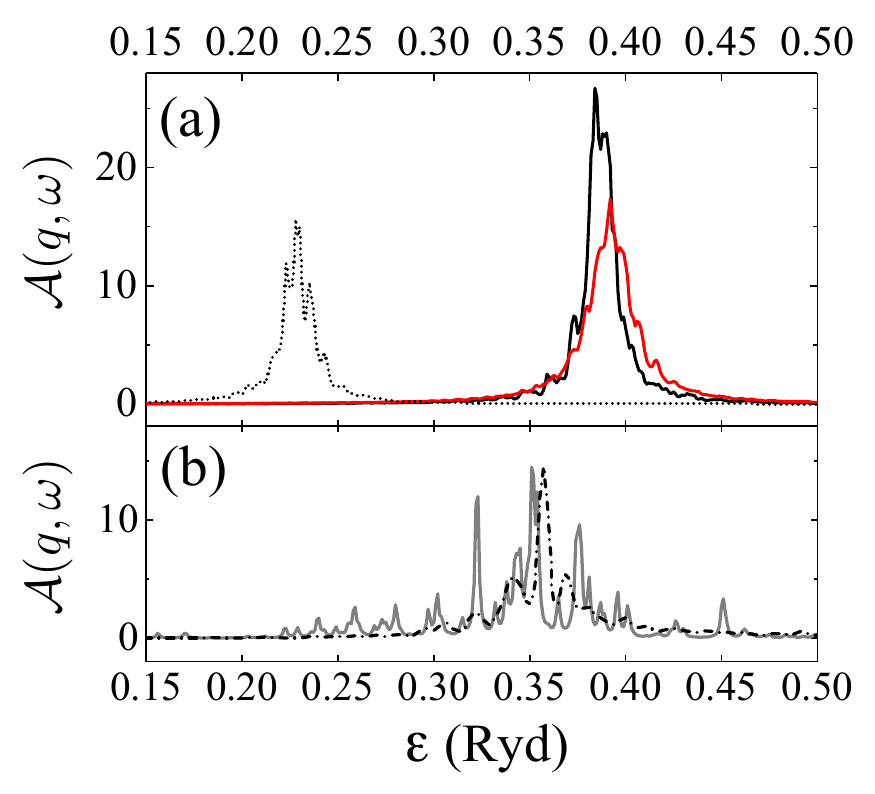}
\caption{Panel (a), optical response function ${\cal A}(q,\omega)$ at $r_s = 3$ and $R = 3.1$ nm, calculated at $T$ = 0 (full black curve) and $T$ = 50000 K (full red curve) in vacuum, and in a surrounding medium with $\epsilon_m = 4$ at $T$ = 0 (black dotted curve). Panel (b), comparison between the present calculation (full grey surve) and that performed using TDDFT for a small sphere containing about 100 electrons, at $T$ = 316 K, in vacuum (black dashed-dotted curve).}
\label{fig2}
\end{figure}

In Fig. \ref{fig2} (a) we show the response ${\cal A}(q,\omega)$ for a sphere ($C = 2$) with radius $R = 3.1$ nm, containing 7445 electrons at $r_s = 3$, in vacuum ($\epsilon_m$ = 1), as a function of temperature from T $= 0$ to T $= 50000$ K. Larger systems can be treated using the present approximation over the same temperature range. We fixed $q = 0.01$ a.u. to approximate the $q \to 0$ limit. We also show the result at T $= 0$ obtained using $\epsilon_m$ = 4, so that the plasmon peak is red-shifted. It is quite interesting to observe that most part of the complex structure visible in the dielectric function (Fig.\ref{fig1}) disappears, and the response results in a rather smooth function with a single peak that can be attributed to the surface plasmon. Secondary structures are visible in a fashion similar to what reported in Ref. \cite{simgas}. For comparison purposes, the response ${\cal A}(q,\omega)$ calculated at $R = 1.34$ nm and T = 316 K is shown in Fig.\ref{fig2} (b) against the companion results from Ref. \cite{simgas}. We observe that the present calculation of the response function provides results very similar to the advanced TDDFT, which represents a reasonable support for the validity of our high temperature calculation. On further increasing the temperature, further smoothing of the peak structure occurs, as expected. Considering that the function ${\cal A}(q,\omega)$ is related to the quantity measured in a real transmission experiment, these results indicate that quantum effects can be difficult to observe in practical conditions.

\begin{figure}
\centering
\includegraphics[width=8.5cm]{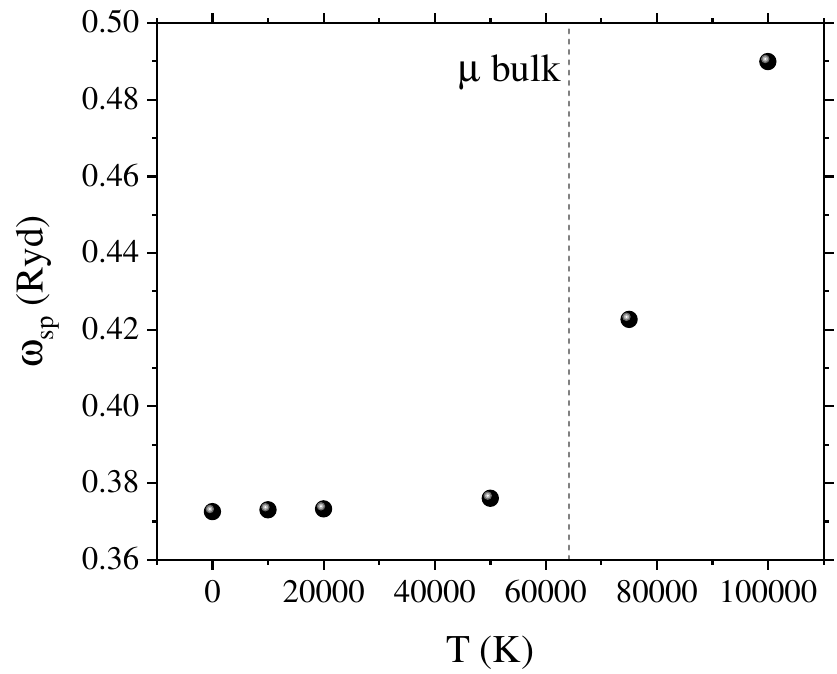}
\caption{Position of the surface plasmon resonance of a gold nanoparticle ($R$ = 1.34 nm, $r_s$ = 3), estimated as the center of mass of the response function ${\cal A}(q,\omega)$, as a function of temperature in the range 0-10$^5$ K. The vertical dashed line corresponds to the Fermi temperature in bulk gold.}
\label{fig3}
\end{figure}

Fig. \ref{fig3} shows the trend of the surface plasmon resonance position as a function of temperature. A continuous increase of the energy of the resonance is readily observed in the range from 0 to $10^5$ K, by a quantity that is well detectable with the current experimental capabilities. This result suggests that a pump and probe experiment with adequate time resolution is feasible. Therefore, by heating small metal particles by means of a short (say 50 fs) pump laser pulse, the electronic temperature could be determined by measuring the position of the surface plasmon resonance with a probe delayed by a time interval short enough to maintain a hot electron plasma and no "much" energy transferred to the lattice.

\section{Conclusions}

The present results show that a free electron approximation provides good results for the description of the dielectric function of electron gas fragments, particularly when the fragment is rather large and contains thousands of electrons, at thermal energies comparable to or higher than the Fermi energy. The proposed approach is quite flexible and it is easy to include the exchange-energy contribution. Further improvements can also be considered. Considering that the dielectric function (Eq. \ref{eq1}) is the sum of several state-dependent contributions, the state populations can be changed by properly modeling the effect of an incoming photon beam that produces transitions from one state to the other. This also contributes to the increase of the sample electronic temperature or to the presence of hot carriers \cite{rev16}.

\section{Acknoledgements}

The Sincrotrone Trieste SCpA is acknowledged for the support the PhD program in Physics. This research has been partially supported by the PRIN (2012Z3N9R9) project of MIUR, Italy.

\end{document}